\pdfoutput=1

\documentclass[pre,twocolumn,showpacs,nofootinbib,amsmath,amssymb]{revtex4-1}

\usepackage{graphicx}

\begin{document}

\title{What do vote distributions reveal? } 
\author{Angelo Mondaini Calv\~ao$^1$}
\thanks{E-mail: oangelo@gmail.com}
\author{Nuno Crokidakis$^{1,2}$}
\thanks{E-mail: nuno@if.uff.br}
\author{Celia Anteneodo$^{1,3}$}
\thanks{E-mail: celia.fis@puc-rio.br}

\affiliation{
$^{1}$Department of Physics, PUC-Rio, Rio de Janeiro, RJ, Brazil \\
$^2$Institute of Physics, Universidade Federal Fluminense, Niter\'oi, RJ, Brazil\\
$^{3}$National Institute of Science and Technology for Complex Systems, Rio de Janeiro, Brazil}

\date{\today}

\begin{abstract}
\noindent
Elections, specially  in countries such as Brazil with an electorate of the order of 100 million people, 
yield large-scale data-sets embodying valuable information on the dynamics 
through which individuals influence each other and make choices. 
In this work we perform an extensive analysis of data sets available for Brazilian proportional elections of
legislators and city councillors  throughout  the period 1970-2012, which  embraces two distinct political regimes: 
a military dictatorship and a democratic phase.
Through the distribution $P(v)$ of the number of candidates receiving $v$ votes,   
we perform a comparative analysis of different elections in the same calendar and as a function of time. 
The distributions $P(v)$ present a scale-free regime with a power-law exponent $\alpha$  
which is not universal and appears to be characteristic of the electorate. 
Moreover, we observe that $\alpha$ typically increases with time. 
We propose a multi-species model consisting in a system of nonlinear differential equations 
with stochastic parameters that allows to understand the empirical observations. 
We conclude that the power-law exponent $\alpha$ constitutes a measure of the degree of feedback of the 
electorate interactions. To know  the interactivity of the population is relevant beyond 
the context of elections, since a similar feedback may occur in other social contagion processes. 
\end{abstract}



\maketitle

\section{Introduction}

In the last few decades, special attention has sparked  amongst statistical physicists the study of  
elections~\cite{castellano,galam_book,galam_review,sznajd_app,hans}. 
An election can be seen as a large scale event that provides a huge amount of real data about 
 people choices resulting from  collective processes. 
Then,  its  analysis may 
reveal important  hints on how people are influenced and opinions propagate. 
The number of data is particularly huge in countries such as Brazil,  with a large and diversified electorate. 
Moreover, differently from the case of presidential or  governor elections, with few candidates, 
in an election  for legislators or for city councillors,  there is also a large number of candidates, 
of the order of a few thousands in the whole country. 
This provides a wide spectrum of choices that allow to probe how people preferences distribute. 

Plausibly following these motivations, pioneering   works analyzed the Brazilian proportional elections 
of 1998~\cite{raimundo1,bernardes1,bernardes2,lyra,gonzalez},  
for federal and state deputies   in the most populated states  S\~ao Paulo and Minas Gerais.
It was first claimed that the probability density function (PDF)  
of the number of votes $P(v)$ for deputies  presented, 
as a common feature, a power law decay $P(v)\sim 1/v^\alpha$, with exponent 
$\alpha=1$~\cite{raimundo1,bernardes1,bernardes2,lyra,gonzalez,raimundo2,prado}. 
Diverse opinion dynamics models were then proposed 
to explain such behavior~\cite{bernardes1,bernardes2,gonzalez,fontoura}. 
The simple Sznajd dynamics (only agreeing pairs of individuals can convince their neighbors~\cite{sznajd_app})
appeared to be enough to explain the power-law exponent $\alpha=1$. 
A robust result  almost independent  of the network  properties~\cite{bernardes1,bernardes2}.  
A  simple contagion model~\cite{fontoura} was also able 
to reproduce the $1/v$ behavior, but the small-world effect appeared to be crucial in that case.  

For Brazilian city councillors, the exponent was found to be larger than unit~\cite{lyra}.
Elections with proportional rules in other countries were also analyzed, 
like  India~\cite{gonzalez},  Finland~\cite{raimundo2} and  Italy~\cite{fortunato},
revealing that the power-law exponent, or even the power-law itself, was not robust~\cite{fortunato}. 
As a consequence, it was argued that an essential feature to capture a universal behavior 
was to take into account the role of parties~\cite{fortunato,raimundo2,chatterjee}.  
In fact, it is sound that a voter  chooses first a party, following its ideology,  
and then  the voter chooses a candidate belonging to that party. 
Moreover, voters may be influenced by the fact that  the  total amount of votes for a party  
will decide the corresponding number of seats. 
However, at least in the Brazilian elections, 
several  features hamper a meaningful normalization by party:    
i)~electoral choices largely rely on the personal characteristics of candidates~\cite{tese} and 
although voting for a party is allowed, most people vote mainly for candidates directly 
(within an open list system),  
ii) there are over 30 political parties, distributed in a broad spectrum of orientations~\cite{wiki_list} 
but parties occasionally merge together or split and there are also alliances, 
in some cases of mixed political orientation, and 
iii) there are no electoral  thresholds to disqualify  parties with low representativity.  
Then,  we will not tackle here  aspects related to parties. 
We aim to gain insights on the formation of preferences of the electorate about candidates only.

We analyze  Brazilian elections  for legislators  during the period 1970-2010, 
for which data are available at the website of the Brazilian Federal Electoral Court~\cite{TSE}. 
This period encompasses  elections that took place during the civilian governments after 1986, 
as well as during the precedent military dictatorship, allowing to investigate the 
impact of very different political regimes on  vote distributions. 
Complementarily, we also consider the elections for city councillors during 2000-2012.  
A detailed description of the electoral context is provided in the Appendix.
%
After the empirical analysis presented in Sec.~\ref{sec:analysis}, 
we will introduce in Sec.~\ref{sec:model} a model that allows to understand  the  
features observed in real data analysis.
A comparative study of model and empirical data follows in Sec. \ref{sec:comparison}, 
ending by a discussion in Sec.   \ref{sec:final}.

\section{Vote distributions}
\label{sec:analysis}

\begin{figure}[b!]
\includegraphics[width=0.5\textwidth,angle=0]{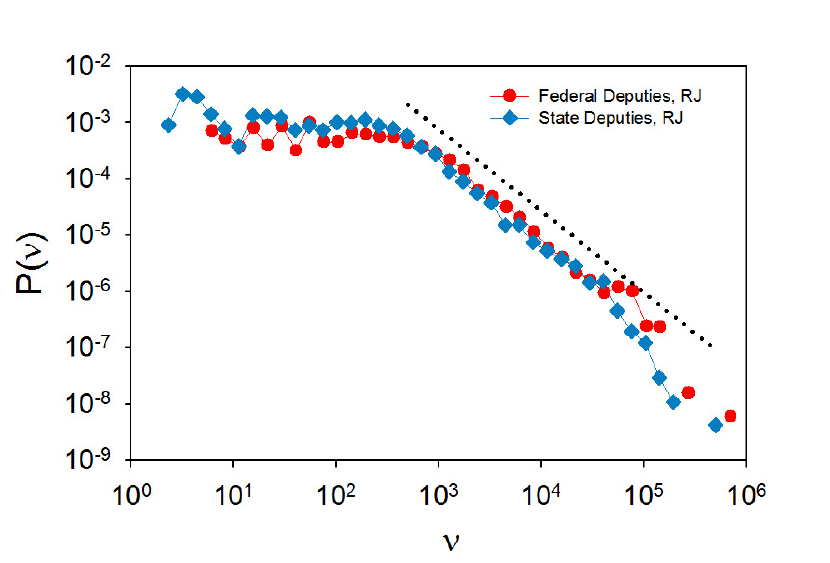}
\caption{Probability density of the quantity of votes $P(v)$ received by candidates  
 for  deputies of Rio de Janeiro state in 2010.   Data are logarithmically binned. 
In this and other figures, thin lines are a guide to the eyes.  
The  straight line with slope -1.45  is  drawn for comparison. 
}
\label{fig:rio}
\end{figure}

The typical shape of the probability density function $P(v)$ of the quantity of valid votes  received by candidates 
is illustrated in Fig.~\ref{fig:rio}, by means of the data of the two elections for deputies of  
 Rio de Janeiro state in 2010. 
Both PDFs  display   an initial almost flat  region, within statistical fluctuations, 
and a power-law regime over two decades, 
truncated by a rapid decay. 
That is, relatively small, intermediate and large numbers of votes show  
different statistics.   
Small numbers of votes (up to  $\sim 10^3$ from a total of  $\sim 10^7$, in the case of the figure) 
are almost uniformly distributed, intermediate numbers (approx. in the range $10^3-10^5$) 
are scale-free distributed~\cite{barabasi} and very large quantities of votes are outliers,  
typically corresponding to the number of votes received by famous people or very popular experienced politicians. 
Actually they are the election winners!  
Although the crossover between the uniform and scale-free regimes, as well as the exponent $\alpha$, may  
change from one case to another, the profile of $P(v)$ shown in Fig.~\ref{fig:rio} displays 
the general form found for other electoral years  and for other states.
Moreover, notice that both PDFs shown in the figure are very close to each other, specially the scale-free region with  
power-law exponents that coincide within error bars~\cite{errors}. 
The main difference is in the  level of the flat region, indicating a 
 larger fraction of candidates receiving few votes in the elections for state deputies, which also 
present a  larger number of candidates (see the Appendix).

\begin{figure}[b!]
 \hspace*{-4cm} (a)\\
\includegraphics[width=0.4\textwidth,angle=0]{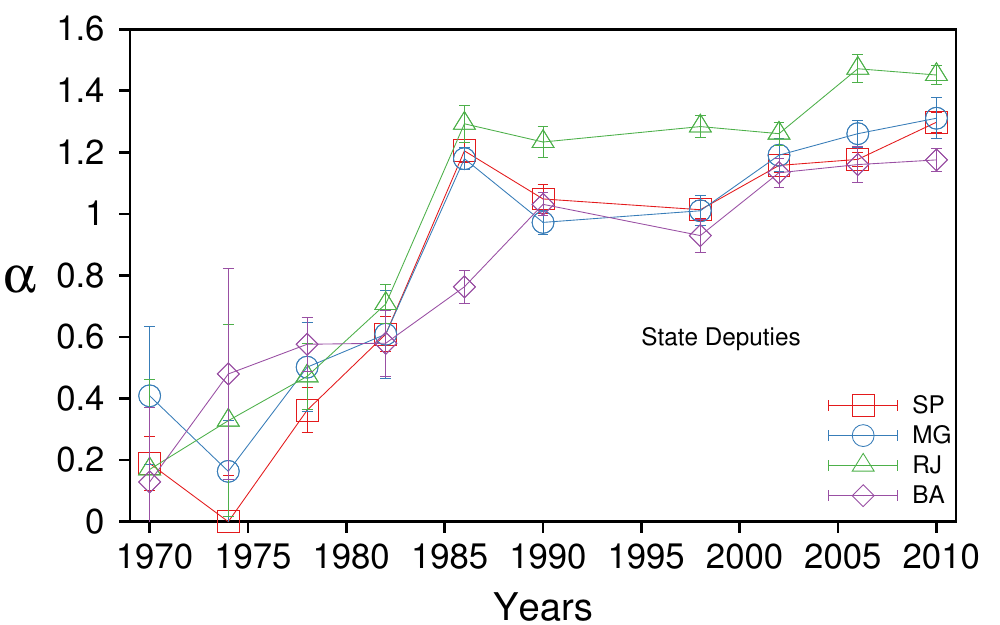}\\
 \hspace*{-4cm}(b)\\
\includegraphics[width=0.4\textwidth,angle=0]{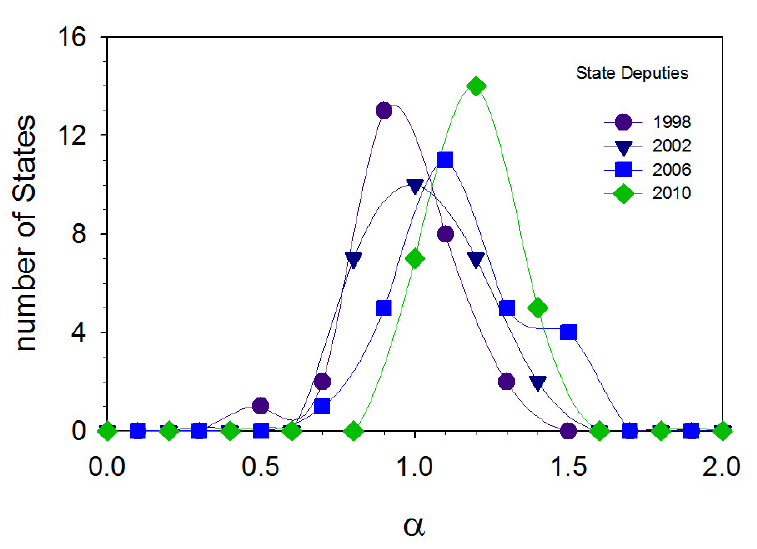}\\
 \hspace*{-4cm}(c)\\ 
\includegraphics[width=0.4\textwidth,angle=0]{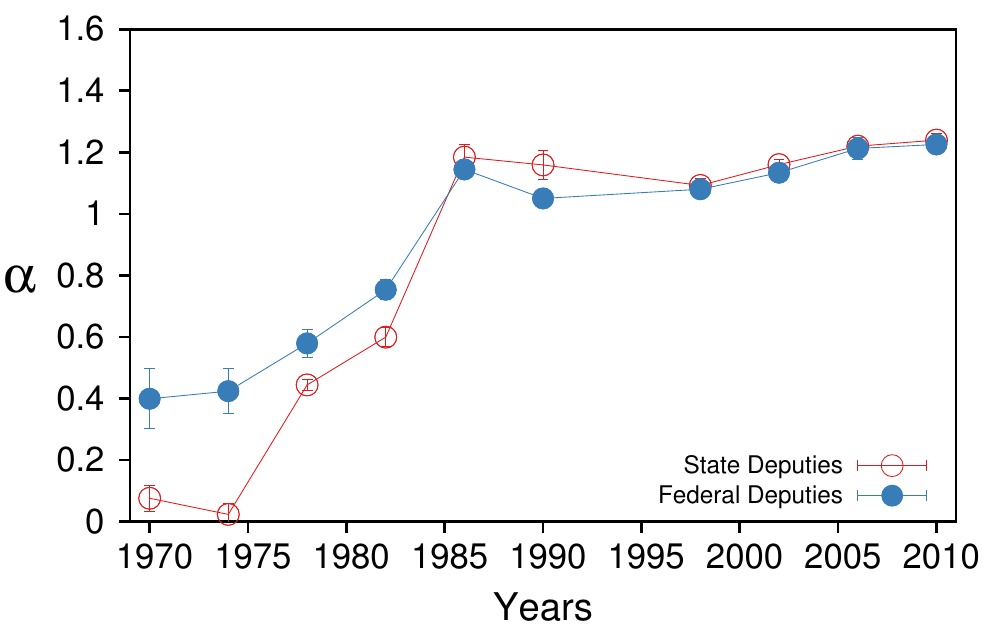}
\caption{ (a) Exponent  $\alpha$  of the PDF of votes $P(v)$ as a function of the electoral year,  
from elections for   state deputies, in the four most populated 
states: S\~ao Paulo (SP), Minas Gerais (MG), Rio de Janeiro (RJ) and Bahia (BA). 
(b) Histogram of $\alpha$  across all the states for the years indicated on the figure. 
(c) Exponent $\alpha$ as a function of time, for both deputies and whole Brazilian electorate. 
A non-linear least squares  fit was used to obtain $\alpha$ from the 
intermediate scale-free region. 
}
\label{fig:alfa}
\end{figure}

For each PDF, the exponent $\alpha$ was obtained by fitting a power-law to the scale-free region. 
Figure~\ref{fig:alfa}a shows $\alpha$ as a function of the electoral year 
for the state deputies elections in the four most populated states. 
The outcomes for federal deputies in these states, as well as the results for both deputies 
in the remaining states, display similar tendencies. 
In contrast with the results   for S\~ao Paulo and Minas Gerais in 1998~\cite{raimundo1,lyra}, 
the exponent $\alpha$ typically differs from 1. 
The histogram of the values of $\alpha$ obtained for each state is shown in Fig.~\ref{fig:alfa}b 
for the case of state deputies.
A comparison of the values of $\alpha$ as a function of time for both deputies elections, 
grouping the whole Brazilian electorate, is shown in Fig.~\ref{fig:alfa}c.
Notice that, specially in the democratic period, the values of $\alpha$ for both elections at the same 
calendar coincide within error bars~\cite{errors}. 

For each state, as well as for the whole Brazilian electorate, 
one observes   that   $\alpha$ typically increases with time, except for the peak at 1986. 
The two political regimes clearly have a different impact in the behavior of parameter $\alpha$. 
A rapid increase with time occurs during the dictatorship and a slow one during democracy.
The variability of the exponent $\alpha$ cannot be explained simply by differences 
in the quantity of votes $N_v$ or of candidates $N_c$, although all tend to increase with time. 
In fact, for instance, the most notable  changes in $\alpha$
with a record value in  1986, occur in a period of relative stability of  the electorate size. 
On the other hand, for the full Brazilian electorate, 
it is clear in Fig.~\ref{fig:alfa}c  the proximity of the values of $\alpha$ for federal and state deputies  in the 
democratic period,   despite the number of candidates $N_c$ differs (see Appendix), 
and mainly despite the pools of candidates are distinct.

\begin{figure}[h!]
\includegraphics[width=0.45\textwidth,angle=0]{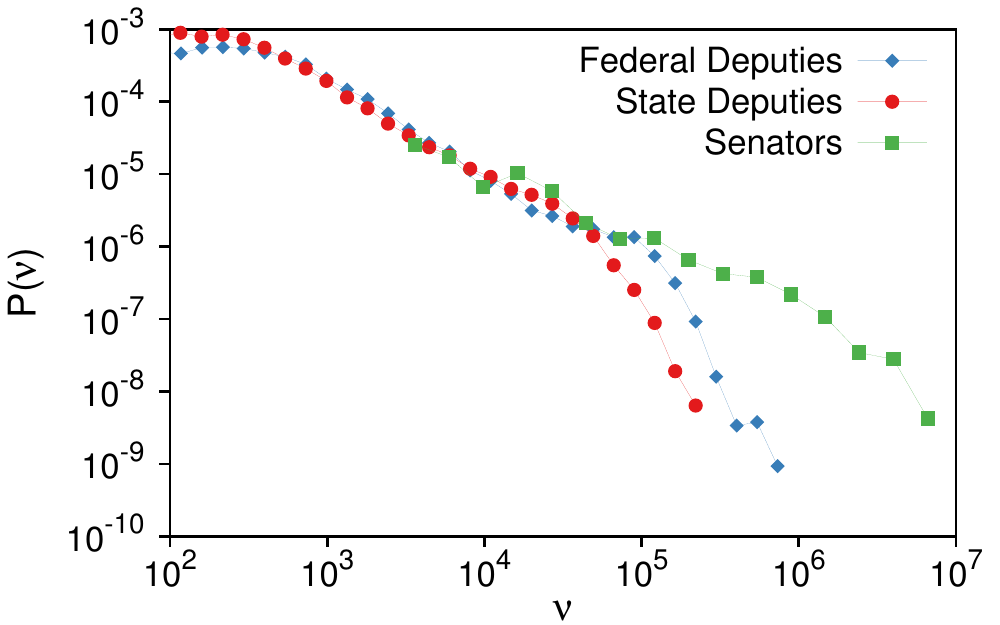}
\caption{Comparison of the PDFs of votes for deputies and senators in 2010, for the whole 
Brazilian electorate.  
Thin lines are a guide to the eyes.
}
\label{fig:senat}
\end{figure}

For senators, the  number of candidates per state is too small  (e.g., only eleven in Rio de Janeiro in 2010) 
to build an histogram, however we  obtained  $P(v)$ for the whole electorate.
$P(v)$ also decays as a power-law, although the plateau phase is lacking, 
reflecting the absence of candidates with few votes. 
In Fig.~\ref{fig:senat}, we compare the PDFs of votes for senators and deputies for the whole Brazilian electorate in 2010. 
Notice the coincidence of the PDFs in the scale-free region, 
for the three  different elections with different candidates but the same electorate.

The present statistical analysis indicates that:  
i)   $\alpha$ is not universal, ii)  it tends to increase with time, 
iii)  for a given electorate, it takes a defined value. 
That is, $\alpha$ seems to be associated to  some property of the electorate or its 
interactions, independently of the pool of candidates and beyond the electorate size. 
Therefore, the scale-free regime may reflect a  fundamental feature 
of the complex collective processes involved.

\section{Model}
\label{sec:model}

Previously proposed models reproduce the $1/v$ law but do not allow to describe 
the PDF  decay of the form $1/v^\alpha$ with generic $\alpha$ that we observe in real data. 
We propose a simple model, where we describe the temporal evolution 
of the quantity of votes $v_i(t)$ received by each candidate $i$ as a continuous variable. 
This is justified by the fact that 
the fractions $v_i/N_v$  tend  to assume continuous values in the limit of a very large number of voters. 
In analogy to multi-species models of population dynamics, we consider 
that the number of voters for  candidate $i$, hence  the quantity of votes $v_i$,  
follows a logistic-like  growth dynamics. 
Then, the time evolution is governed by a set of  ordinary differential equations of the form
\begin{equation} \label{model0}
\frac{dv_i}{dt} =  r_i ( v_i+c)^\gamma f_i\left( \{ v_j\} \right) \,,
\end{equation}
for $1\le i, \, j \le N_c$, where $r_i$, $c$ and $\gamma$ are positive parameters,  and 
$0\le f_i \le 1$  limits the quantity of votes.

Randomness is a realistic ingredient of social dynamics,  
relative both to individual attitudes and to social influences~\cite{castellano}. 
Then,  one should consider in principle the equation parameters   as stochastic variables to reflect 
the heterogeneity of the electorate and its interactions. 
However, for simplicity, we will  assume heterogeneity in the parameters where it is more crucial, as discussed below.

The coefficient  $r_i$  represents a kind of fitness of  candidate $i$. 
It is  related to the capacity of persuasion  determined by a set of 
attributes attached to the candidate, such as its political proposal or personal appeal. 
A negative value would lead to a vanishing fraction of votes. 
Then, for the candidates that win votes, the parameter $r_i$ should take positive values. 
We consider that  the population of candidates is heterogeneous with respect to 
this parameter, according to a given PDF $P_r(r)$.
A similar kind of heterogeneity has been assumed in the cellular automata of 
Ref.~\cite{bernardes2}, where the probability of convincing is different for each candidate in a first 
stage where only candidates can influence voters.

The propagation of opinions about  candidates is a sort of branching or multiplicative process~\cite{lyra,raimundo1}, 
in which a supporter of a given candidate can persuade, with a given probability, a fraction of its neighbors, 
and each one of them in turn can influence others. 
In the standard case,  the intrinsic rate of growth per capita of the number of votes $v_i$ is constant, given by $r_i$. 
However, if  there were a positive feedback between individuals,   the rate per capita would increase  
with $v_i$, which can be   described by $\gamma>1$. 
A similar idea of self-reinforcing mechanisms, that can lead to herding behavior, have already been considered 
in the context of financial bubbles~\cite{sornette}. 
But avalanches of opinion propagation can occur in other contexts too~\cite{watts}.
The ``better connected'' is the electorate, the larger $\gamma$.  
This does not mean higher mean connectivity or higher probability of contagion, but that the probability of 
contagion becomes larger as the number of followers increases. 
A negative feedback may also occur and is represented by $\gamma<1$. 
It is true that one could associate a different $\gamma$ to each 
candidate, reflecting the particular feedback of the community 
in which the candidate exerts influence. 
However,  we will consider that $\gamma$ is predominantly homogeneous across a given electorate.

The introduction of parameter $c$ allows to contemplate the existence of two realistic regimes,  
as considered in a previous model of elections \cite{bernardes2}.  
In an initial phase when $v_i \ll c$,  candidates influence voters either directly or through their staff, 
independently of the number of followers of each candidate. 
Otherwise,  when the number of followers becomes large enough, people interact and 
become influenced by other electors too. 
Then $c$ can be related to the spreading which is independent  of the number of followers $v_i$, 
and may  also include the effect of the media   
through which a candidate can spread its influence and gain voters.  
This parameter could be in principle a random parameter, different for each candidate, 
but also in this case  we will take it as homogeneous.

Finally,  we consider a logistic factor of the general form 
\begin{equation} \label{logistic}
f_i (\{ v_i\}) =   1- \frac{v_i}{\beta_i N_v} - \frac{\sum_{j\neq i} \beta_{ij}v_j}{N_v}  \,,
\end{equation}
 that assumes that each quantity of votes $v_i$ follows a logistic growth, 
up to   a   maximum  value $\beta_i N_v$, with $0<\beta_i\le1$, 
corresponding to the portion of the electorate a candidate would conquer in the absence of other candidates. 
The last term is responsible for coupling the equations and describes  
the inter-specific competition amongst candidates in their struggle for conquering voters, 
where $\beta_{ij}$ measures the competitive effect of candidate $j$ on candidate $i$.  
In modeling voting processes, it is commonly considered as a more realistic situation, 
that  when the election occurs the opinion dynamics has not necessarily attained yet the steady state~\cite{bernardes2}. 
Then, as soon as the logistic term affects only the long-time dynamics, 
for simplicity, we consider that $\beta_{ij}=\beta_i=1$,   $\forall i,j$. 

Besides randomness in the equation parameters, 
another source  of variability may reside in the initial distribution of votes $v_{0i}\equiv v_i(0)$. 
It is clear that there are candidates with a political history and hence may start 
with a given community of followers, then, it would be interesting to investigate  realistic 
initial distributions. 
However,  if the initial numbers of votes are relatively small, as expected for the majority of the candidates, 
their precise values will not affect the distribution at later times, at least for  $v_i \gg 1$. 
Then we will set  $v_{0i}=1$  for all candidate $i$, corresponding to the minimal value of its own vote.

\begin{figure}[h!]
\hspace*{4mm}
\includegraphics[scale=0.95]{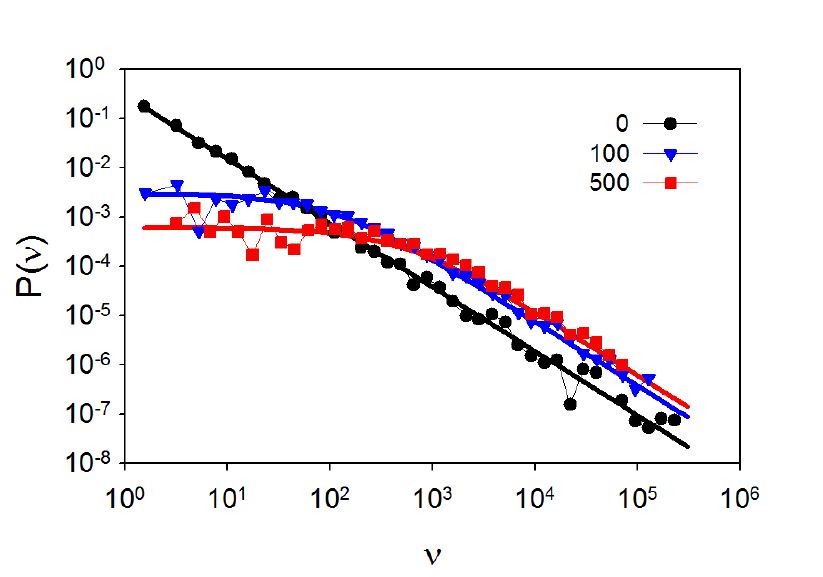}\\ 
\caption{$P(v)$ from model simulations  of the set of Eqs. (\ref{model}), with 
$\gamma=1.3$ and $r_i$ uniformly distributed in the real interval $[0,1]$, 
 for    different values of $c$ indicated on the figure, with $N_v=10^7$ and $N_c=10^3$, 
in a single realization. 
The initial condition was $v_i(0)=1$ for all $i$, 
and $P(v)$ was computed from the values of $v_i$ at the final steady state. 
The thick lines  correspond  to the theoretical prediction given by Eq. (\ref{teo1}). 
For $c=0$ the pure power-law $1/v^{1.3}$ arises.  
}
\label{fig:model}
\end{figure}

Summarizing the model description,  
under the above assumptions,  the evolution equations are reduced to the simple form
\begin{equation} \label{model}
\frac{dv_i}{dt} =  r_i (v_i+c)^\gamma \bigl( 1- \sum_{j}  v_j/N_v  \bigr) \,,
\end{equation}
for $1\le i \le N_c$, where $\gamma$ and $c$ are positive constants and  $r_i$ is the only random 
parameter. We assume that it is uniformly distributed in $[0,1]$. 
(Notice that introducing a different maximal value of $r$ will be equivalent to rescaling the time, then we
chose the unit interval.)
That is, the model depends on only two parameters that appear to be the most relevant ones. 
From the numerical integration of this set of differential equations, we obtained $P(v)$. 
We considered the values of $v_i$ at the steady state, however the results of the model 
are not significantly affected if we do not wait for the steady state to be reached.
 As illustrated in Fig. \ref{fig:model}, 
the model produces outcomes qualitatively similar to the empirical ones, with a flat region and 
a power-law decay.
Notice however that the last phase of rapid decay of the distribution is not present. 
In fact, as discussed above, very popular candidates gaining  very large numbers of votes are those 
contributing to the cutoff, but they are not contemplated by our model in its present form, 
which assumes a uniform distribution of $r$. 
This could be readily improved by modifying  $P(r)$, to take  into account people with outstanding fitness,  
but we  are interested here in the free-scale regime.

\begin{figure}[h!]
\includegraphics[width=0.4\textwidth,angle=0]{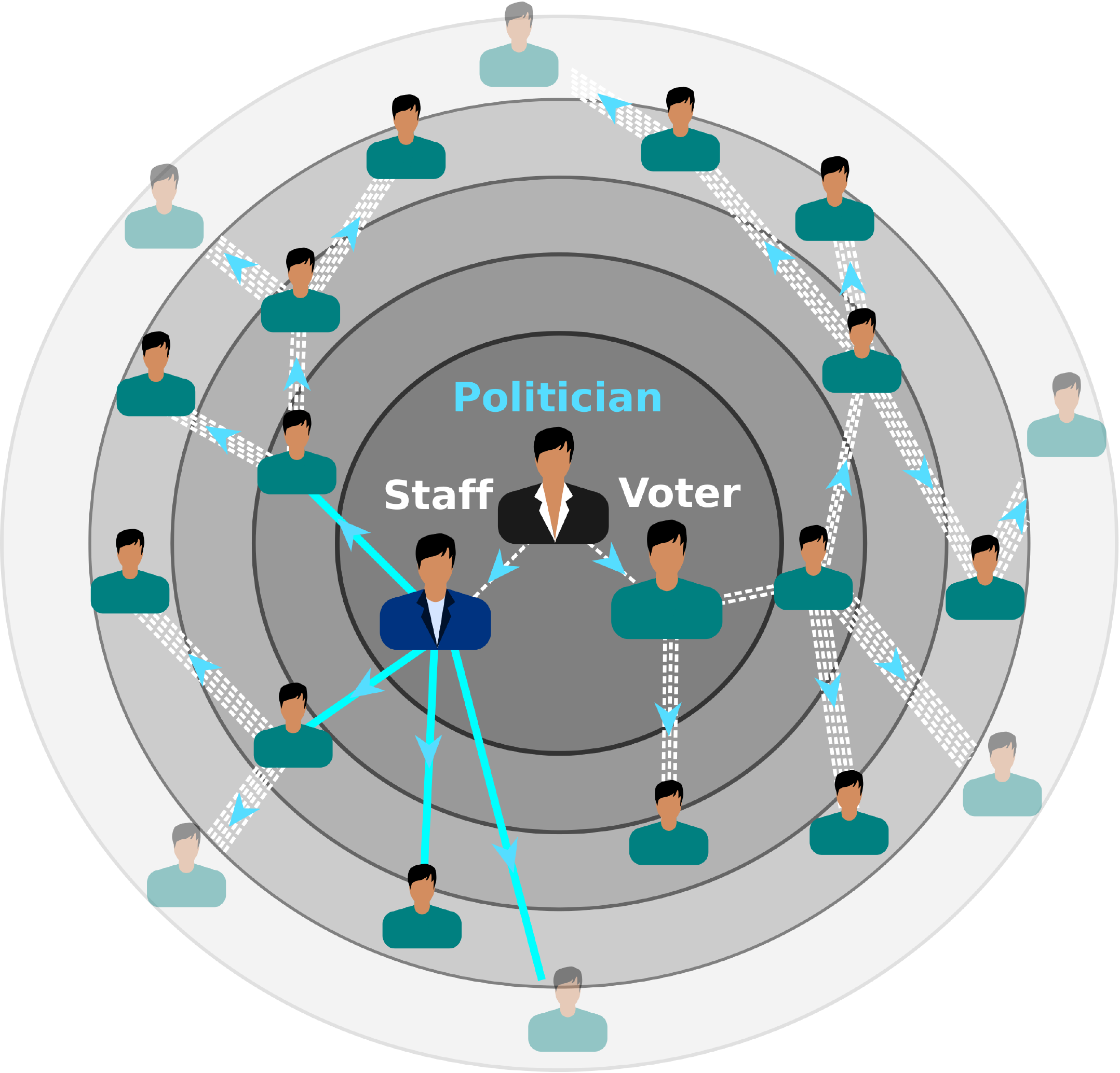}
\caption{Pictorial representation of the model. A candidate gains followers by means of 
either its interaction (or that of its staff) with the electorate or  by means of a 
cooperative interaction of the electors.  
The arrows indicate contagion of ideas in time (evolving radially outwards).  
The figure illustrates the case of positive feedback: the probability of contagion 
increases (represented by thicker arrows) as the population of followers increases.  
}
\label{fig:picture}
\end{figure}

A pictorial representation of the model is given in Fig.~\ref{fig:picture}.
It illustrates the  two mechanisms of dissemination of ideas of a candidate $i$: 
i) without participation of the electorate, for low $v_i$ compared to $c$ 
the propaganda of the candidate dominates the diffusion process,  
and ii)  when the number of followers  $v_i$ is large enough, the phase of interactivity of the electorate occurs with 
 feedback ($\gamma \neq 1$)  or  without it ($\gamma=1$).  
  The (positive) feedback is represented in the picture by thicker lines. 
	In the measure than the number of followers of a candidate grows, 
	the probability or rate per capita of convincing the nearest neighbors 
	in the network of contacts also increases.
The scale free regime is associated precisely to this interactive phase. 
Actually, there is a third regime,  when the number of undecided  people becomes small 
($\sum v_i(t)\sim N_v$, hence the factor $f_i$ tends to zero), 
a phase of competition between candidates operates, that is not represented in the picture. 
%

The distribution $P(v)$ resulting from the model can be analytically evaluated. 
At sufficiently short times the evolution of  $N_i$ will not feel   the competition term and 
the equations are uncoupled. 
This corresponds to an initial phase in which  voters interact mainly with candidates 
in its network of influence.
As a first approximation, integrating Eq.~(\ref{model}) with $f_i=1$ for all $i$, we have 
\begin{equation} \label{approx}
v_i(t) \simeq v_{i0} \left(1 - (\gamma-1) r_i\, t\, v_{i0}^{\gamma-1}   \right)^{-1/(\gamma-1)} \,.
\end{equation}
This approximate expression is expected to hold up to a time close to the upper threshold given by 
the condition $\sum v_i(t)=N_v$ ($f_i=0)$. 
Eq.~(\ref{approx}) provides the dependence between the random variables $v$ and $r$ that allows to 
find the relation between their PDFs, by equating $P(v)dv=P_r(r)dr$. 
In particular, let us consider that   $ r $ is uniformly distributed in the interval $[0,1]$. 
In this case, we obtain that the distribution of $v(t)$ is 
\begin{equation} \label{teo1}
P(v) =   \frac{\cal N}{ (c+v)^\gamma}\,,
\end{equation}
for $1 \le v \le v_{max}$, where the normalization factor is ${\cal N} \simeq (\gamma-1) (c+1)^{\gamma-1}$.  
In fact this approximate expression is in good accord with the results of the numerical integration of 
the model (see Fig.~\ref{fig:model}). 
It also satisfactorily describes real distributions, 
 with only two parameters $c$ and $\gamma$, which control the crossover, between 
the plateau and the scale-free regimes, and the power-law decay, respectively. 
Then Eq.~(\ref{teo1})  allows to identify the fitting exponent $\alpha$ 
with the intrinsic exponent $\gamma$, furnishing a direct interpretation for the origin of the scale-free regime. 
Namely,   the exponent $\alpha \equiv \gamma$   
can be associated  to the degree of feedback of the spreading processes.

\section{Comparisons with real data}
\label{sec:comparison}

Let us first analyze size effects. 
In all the empirical cases the electorate size, measured by $N_v$, 
is very large compared to  the number of candidates $N_c$. 
Within the realistic range in which $N_v \gg N_c$ holds, 
the  effect of an increase of $N_v$ on the PDF of votes (not shown)  is to extend the power law range, as 
soon as larger values of $v$ become possible. 
As a consequence, the flat level decreases to preserve the norm. 
This is the same effect observed in real data (Figs.~\ref{fig:rio} and \ref{fig:senat}).
The power law range also increases if $N_c$ decreases (as illustrated by the cases of deputies  
in Fig.~\ref{fig:senat}).

\begin{figure}[t!]
 \hspace*{-4cm} (a)\\
\includegraphics[scale=0.75]{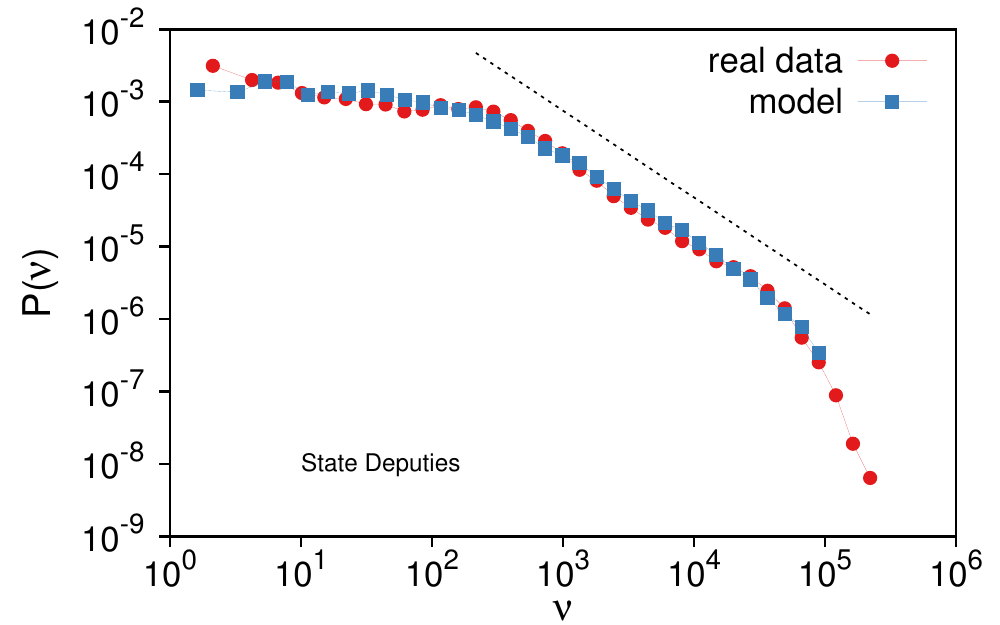} \\
\hspace*{-4cm} (b)\\
\includegraphics[scale=0.75]{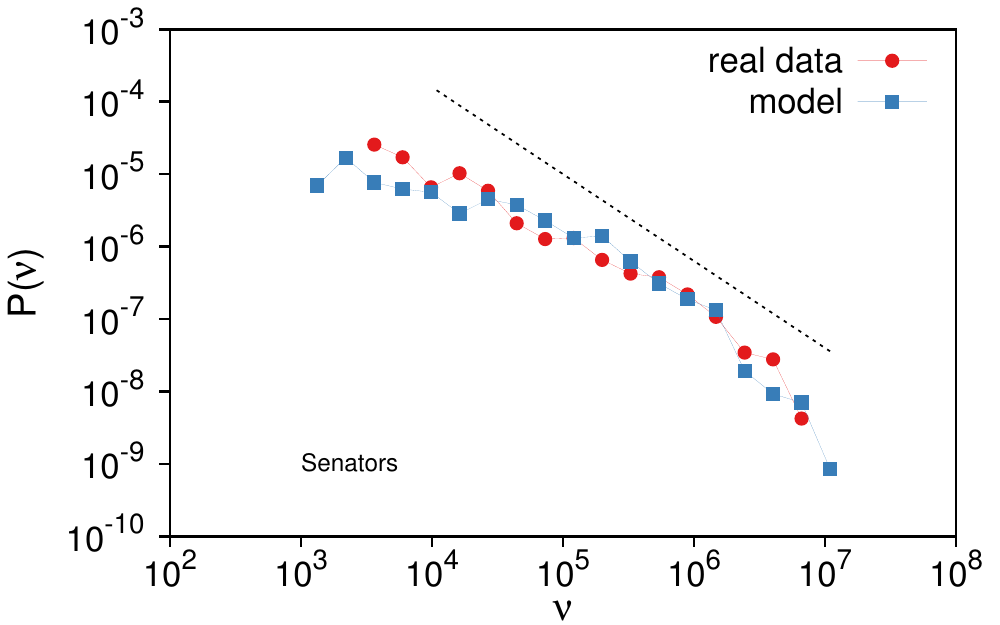} 
\caption{Comparison of the PDFs from  model simulations and real data, for state deputies (a) and senators (b),  
with grouped data for the whole country in 2010. 
Simulations were performed with $\gamma=1.2$, that corresponds to the average value in Fig.~\ref{fig:alfa}b,  
 $c=300$ (a) and $c=4\times 10^4$ (b).  
In both simulations we grouped 26 samples with the values of $N_v$ and $N_c$ of each state, as done for the 
real data.   
The initial condition was $v_i(0)=1$ for all $i$.
 The dashed lines with slope -1.2 are drawn for comparison.
}
\label{fig:comparison}
\end{figure}

In Fig.~\ref{fig:comparison},  we show a comparison of the model outcomes 
with results for state deputies and senators (whole electorate) in 2010. 
We used   $\gamma=1.2$, that corresponds to the average value, across states, in that year 
(from the distribution shown in Fig.~\ref{fig:alfa}b), 
and we run  26 samples with the empirical values of $N_c$ and $N_v$ for each state. 

The case of state deputies is shown in Fig.~\ref{fig:comparison}a. 
A very good accord of the PDF from simulations with that of real data is observed. 
The value of $\alpha$ from fits allows to recover the intrinsic value $\gamma=1.2$. 
In general, simulations fail in describing the initial portion, which is not completely flat. 
This could be improved by using a realistic initial distribution of votes, instead of $v_{i0}=1$ for all $i$. 
Because, although this condition does not affect the distribution of large values of $v$, 
it does affect the distribution of $v\sim 1$.
 
The case of senators is shown in Fig.~\ref{fig:comparison}b.
 A small number of candidates, together with a very large value of $c$, 
 needed to fit the data in that case, may distort the intrinsic value ($\gamma$) of the scale-free regime, 
yielding $\alpha<\gamma=1.2$, as shown in Fig. ~\ref{fig:comparison}b. 
This may explain the apparently smaller value of $\alpha$, compared to those of deputies,  
observed in the case of senators (Fig.~\ref{fig:senat}), with few candidates per state.

\begin{figure}[h]
 \hspace*{-4cm} (a)\\
\includegraphics[scale=0.75]{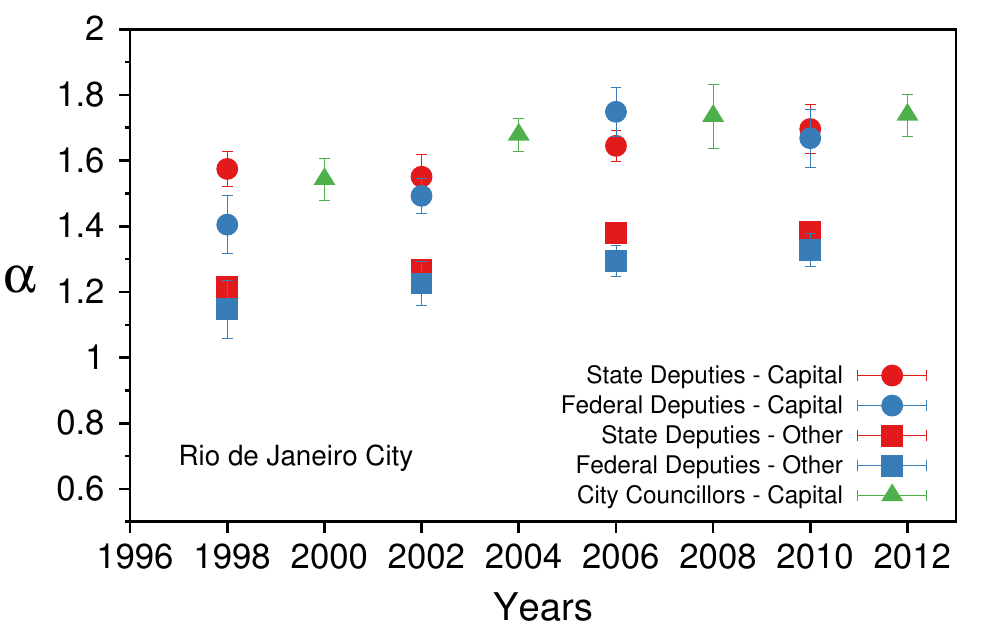}\\
 \hspace*{-4cm} (b)\\
\includegraphics[scale=0.75]{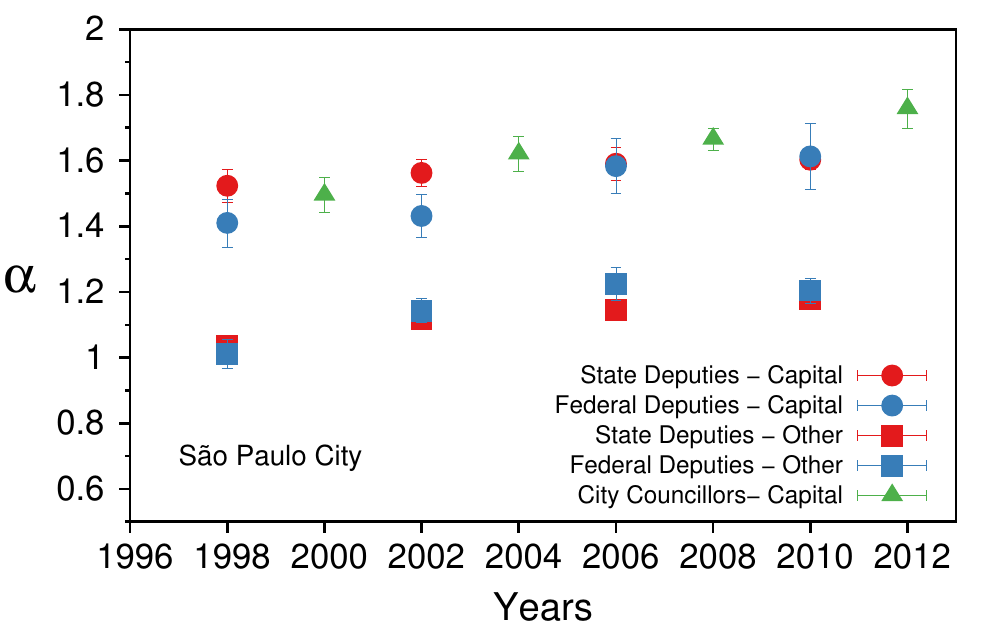}
\caption{ Exponent $\alpha$ of the $P(v)$ for  candidates to deputies (in a given city) and 
city councillors, as a function of time: 
Rio de Janeiro city (a) and S\~ao Paulo city (b). 
The exponent for  deputies in the  remaining of the electorate of the  state, out of the capital (denoted  ``other'') 
is also shown for comparison. 
}
\label{fig:cities}
\end{figure}

Let us consider now the PDFs  for city councillors. 
These PDFs  also present the same characteristic shape shown for legislators,  
but with a larger value of the exponent $\alpha$, as already known~\cite{lyra}. 
Following our model, this exponent might be related to the connectedness or feedback characteristic of each electorate, 
the electorate  of each city in this case. 
If this were true, the same exponent should be observed in other vote distributions for the same electorate. 
To test this hypothesis, we compared distinct elections in the same city. 
In order to do that,  we restricted   the analysis done for legislators to the electorate of each city considered.  
Because of the better statistics, we analyzed only  state capital cities. 
The results for Rio de Janeiro  and S\~ao Paulo cities are presented in Fig.~\ref{fig:cities}. 
Recall that elections for city councillors and deputies occur on alternate even years. 
The exponent $\alpha$ for both deputies coincides within error bars~\cite{errors}.
Notice that although the candidates are different, the points representing the exponent 
$\alpha$ for deputies and city councillors exhibit a continuity, belonging almost to a same curve. 
This reinforces the idea that this exponent is associated to a feature of 
the electorate and its interactions.

We also plotted for comparison the exponent for the remainder of the electorate in each state. 
It displays lower values close to those of the whole state electorate, presented in 
Fig.~\ref{fig:alfa}. 
It is also remarkable that the values of $\alpha$ for large urban conglomerates are larger than 
those for the population   of the corresponding state out of the capital city, 
 consistent with the larger interactivity expected for an urban population.

\section{Discussion}
\label{sec:final}

We analyzed the distributions of votes $P(v)$   
for different Brazilian elections  along the years of 1970-2012.
 $P(v)$ typically presents a flat region followed by a power-law decay. 
The crossover between them may change as well as the power-law exponent, from one 
case to another.  
However, the responses of the same electorate (from a given capital city or state) to 
distinct elections with a different pool of  candidates, 
present very similar exponent $\alpha$
(as observed in Figs. \ref{fig:rio}, \ref{fig:alfa}c, \ref{fig:senat} and \ref{fig:cities}), 
suggesting that this exponent reflects a feature that is predominantly characteristic of the electorate. 
Another noticeable  feature is the tendency of this exponent to increase with   time. 
This tendency occurs not only for global data but also for the data restricted to each state or city, 
  as shown in Fig.~\ref{fig:alfa}.

We introduced a new model of opinion formation in elections, very different from previous ones, 
which consists in a $N_c$-dimensional nonlinear dynamical system, 
similar to multi-species models of population dynamics. 
Stochasticity was introduced through the fitness parameter $r$. 
The model is validated by its ability in describing  the  features of 
the empirical distribution of votes and could be useful to tackle other questions 
related to electoral processes as well.

Under the light of our model, 
the  variability of $\alpha$  appears to be a reflection of  the variability of the mechanisms  
through which voters  interact. 
Namely, the exponent reflects the degree of feedback of the population. 
This explains why the values of $\alpha$ coincide for the same electorate and are larger for urban centers,  
with higher interactivity than rural areas. 
It is also noticeable that typically $\alpha$ is larger than one for the democratic regime, indicating a positive feedback of 
the electorate. 

Contrarily, during the dictatorship, lower than one or even almost null values of $\alpha$ are observed. 
According to the model, they  can be associated   to  negative feedback ($\alpha<1$) 
and  absence  of interaction of the electorate (large $c$), respectively. 
Both are consistent with a dictatorial regime imposing  severe restrictions to social 
interactivity, generating distrust and negative feedback. 
In the measure that  some of the restrictions to democracy gradually relaxed towards the end of the 
military regime,  the increase of $\alpha$ is observed.
Concerning the record value of $\alpha$ in 1986, it may be related to diverse factors that 
favored   the participation of a large and active electorate.  
Besides marking  the end of the military period,
in this opportunity the legislators responsible for the elaboration of 
a new Constitution (the 1988 Constitution) were to be chosen. 
It is remarkable that the singular historical context could in fact have promoted  
a particularly high positive feedback in the Brazilian electorate. 
Besides the intrinsic features discussed above as responsible for a lower value of $\alpha$, 
another factor that may contribute to underestimate the value of $\alpha$ is a low number of candidates, as 
already observed for senators. 
In fact,  candidatures and electoral choices were initially small and 
increased during the transition from bipartidism to pluripartidism.

The increase of $\alpha$ in the last  two decades, during the democratic phase,  
is consistent with the fact that the way people interact is changing. 
The number of people accessing Internet has increased in that period 
and, mainly in the last decade, the number of people connected to social networks 
(such as, Orkut, Twitter, Facebook, etc.) has increased too. 
Moreover, people may participate in more than one of these platforms and  
 members may belong to groups of interest.
These groups of people holding common interests may propitiate a rapid consensus 
among their participants and a positive feedback.  

Our model  could still be improved to be more realistic, 
for instance, by modifying the distributions of $r$  and/or of   the initial distribution of votes $v_0$.  
But, even  in its present form, it captures the more relevant ingredients and it is able to model 
real data. 
In particular, the model allow to relate the scale-free regime 
to the degree of feedback in the electorate interactivity.
The changes of $\alpha$ with time and the larger values of $\alpha$ observed for urban centers 
are consistent with this view.   
It is remarkable that election statistics furnishes a measure, through the effective parameter $\alpha$, 
of the population interactivity feedback which rules the propagation of electoral preferences. 
This measure could be taken into account to design campaign strategies according to the electorate mood. 
But it may be useful also in other contexts since it reflects a property of the population interactivity, which 
governs the propagation of  opinions and influences.

\section*{Acknowledgments}

This work was supported by the Brazilian funding agencies CAPES, CNPq and FAPERJ.

\section*{Appendix: About the data}
\label{sec:data}

We  analyzed  the nominal votes of the elections for different kinds of legislators in Brazil 
(senators,  federal deputies and state deputies) and also city councillors.

Senators and federal deputies are congressmen, representatives of a given state 
(from a total of 26 States in the federation and a federal district),  
in the senate and in the chamber of deputies of the national Congress, respectively. 
While federal deputies are in proportion to the population of the respective State, 
there is a fixed number of senators for any State 
(3 per State, hence a total of 81, including the federal district). 
State deputies are local representatives elected to serve in the unicameral legislature of each state. 
All deputies serve during a 	four-year term, 
while senators serve during  eight-year terms, being renewed 
one-third and two-thirds of the senate  in  alternate electoral calendars. 
All these  elections occur simultaneously  in the same countrywide electoral event, together 
with the election for president and governors, at  four-year intervals. 
From a pool of candidates in each state, an elector can choose  one name of each class 
(two senators, instead of one, in the years when two-thirds are renewed).

Councillors serve during four-year terms in each city council and are voted in the same municipal 
election in which voters chose mayors, every four years, 
in even years alternating with presidential elections.

In the proportional elections here considered, besides the possibility 
of voting on a candidate (nominal vote), 
a citizen can  vote directly on a party (the so called ``legenda'' vote), 
without specifying a particular candidate. 
Actually, only a minority of the electorate practices this possibility (4-18\%). 
Let us also mention that abstention, invalid votes and blank votes   
are all ignored in vote counting for proportional seat allocation purposes.

We scrutinized the distribution of votes for candidates (nominal votes) only. 
That is, blank, invalid votes, as well as, valid legenda votes were not taken into account, 
following our aim of gaining insights on the formation of opinions about candidates. 

\begin{figure}[h!]
\includegraphics[scale=0.8]{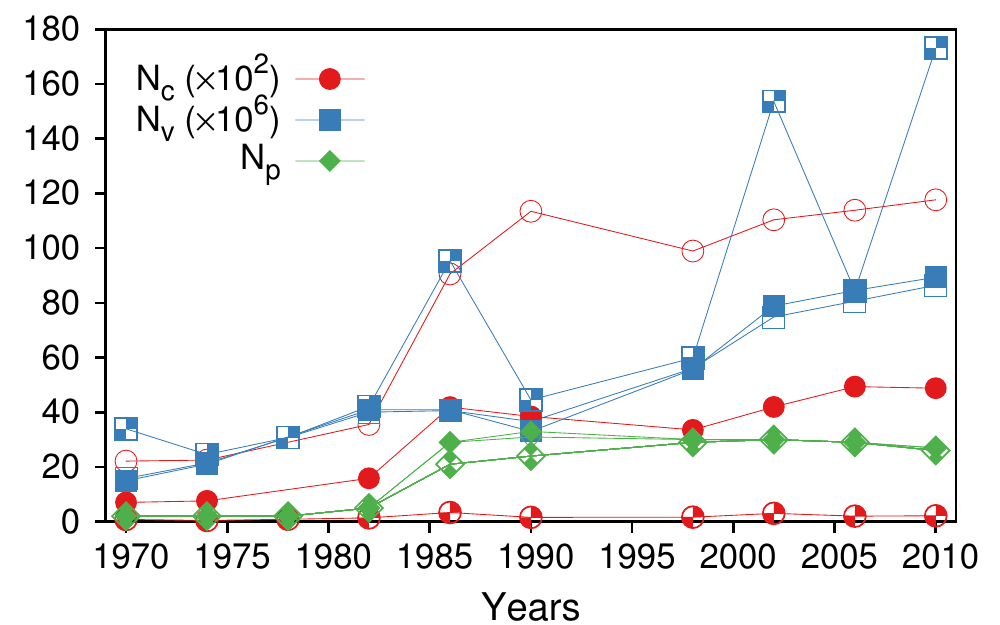}
\caption{Quantity of valid nominal votes $N_v$, number of candidates $N_c$ and number of parties $N_p$ 
for  state (hollow) and federal (filled symbols) deputies, and for senators (half-filled symbols), 
as a function of the electoral year. 
Data for all the Brazilian states were accumulated, including the federal district of Brasilia for senators. 
Data for 1994 were unavailable. }
\label{fig:summary}
\end{figure}

Electoral sizes are shown in Fig.~\ref{fig:summary}. 
The quantity of (valid nominal) votes, $N_v$,  and the number of candidates, $N_c$, 
for deputies and senators of all the states together  vs the electoral year are plotted. 
 
The electorate gradually increased, faster than the population,  
after the establishment in 1985 of the voluntary voting for illiterate   
and minors between sixteen and eighteen years of age.
 
Both kinds of deputies are voted in the same electoral event and only one name 
can be indicated by each voter, then the respective 
total number of voters (hence of votes) $N_v$ are very close. The same occurs for senators, 
except that two names, instead of one, can be elected in alternate calendars 
when two-thirds of the senate is renewed, hence the 
quantity of votes is about twice the quantity of votes for deputies in those occasions.  

The total of candidates $N_c$ for the state legislatures is always larger than for the national one,  
reflecting the larger number of  total  seats to be assigned. 
The number of candidatures for senators is still smaller, not only because the number of 
seats is smaller but also because senators must meet more requirements for eligibility. 
The number of candidates  was lower during the military dictatorship due to the several restrictions, 
increased after the restoration of pluripartidism in 1980 
and attained a relative stabilization at a higher level after 1986.
The number of parties $N_p$ is also represented, 
evincing the  period of bipartidism during the dictatorship and the progressive  
reintroduction of the multiparty system.
 
\begin{thebibliography}{40}
\bibitem{errors} Actually  errors are larger than the bars 
in the figures, because these include only  the error in the fitting parameter due to data fluctuations but do  not include the 
error associated to the selection of the fitting interval.
\end{thebibliography}
\end{document}